\documentclass[conference,a4paper]{IEEEtran}
\usepackage{amssymb}
\usepackage{amsmath,amsthm,amssymb}
\usepackage{graphicx}
\usepackage{color}
\usepackage{algorithm}  
\usepackage{enumerate}
\usepackage[noend]{algorithmic}
\allowdisplaybreaks

\begin{document}
\renewcommand{\IEEEbibitemsep}{0pt plus 2pt}
\makeatletter
\IEEEtriggercmd{\reset@font\normalfont\footnotesize}
\makeatother
\IEEEtriggeratref{1}
\sloppy
\newtheorem{claim}{Claim}
\newtheorem{corollary}{Corollary}
\newtheorem{definition}{Definition}
\newtheorem{example}{Example}
\newtheorem{exercise}{Exercise}
\newtheorem{fact}{Fact}
\newtheorem{lemma}{Lemma}
\newtheorem{note}{Note}
\newtheorem{obs}{Observation}
\newtheorem{problem}{Problem}
\newtheorem{property}{Property}
\newtheorem{proposition}{Proposition}
\newtheorem{question}{Question}
\newtheorem{ru}{Rule}
\newtheorem{solution}{Solution}
\newtheorem{theorem}{Theorem}
\newenvironment{remark}[1]{\textbf{Remark: }}

\def\R{{\rho}}
\newcommand{\A}{{\bf a}}
\newcommand{\bh}{{\bf h}}
\newcommand{\CC}{{\cal C}}
\newcommand{\B}{{\bf b}}
\def\bz{{\bf z}}
\def\bp{{\bf p}}
\newcommand{\I}{{\bf i}}
\newcommand{\p}{{\bf p}}
\newcommand{\q}{{\bf q}}
\newcommand{\x}{{\bf x}}
\newcommand{\y}{{\bf y}}
\newcommand{\z}{{\bf z}}
\newcommand{\M}{{\bf M}}
\newcommand{\N}{{\bf N}}
\newcommand{\X}{{\bf X}}
\def\r {{\bf r}}
\newcommand{\bP}{{\bf P}}
\newcommand{\Q}{{\bf Q}}
\newcommand{\U}{{\bf u}}
\def\d {{\tt d}}
\def\D {{\tt D}}
\def\V {{\tt V}}

\newcommand{\restrict}{{\mathbin{\vert\mkern-0.5mu\grave{}}}}

\newcommand{\glb}{{\tt glb}}
\newcommand{\lub}{{\tt lub}}

\newcommand{\LL}{{\cal L}}
\newcommand{\PP}{{\cal P}}
\newcommand{\pv}{{{\bf p}=(p_1,\ldots , p_n)}}

\newcommand{\qv}{{{\bf q}=(q_1,\ldots , q_n)}}
\newcommand{\Inf}{{\bf t}}
\newcommand{\Sup}{{\bf s}}
\newcommand{\inff}{{\tt t}}
\newcommand{\supp}{{\tt s}}
\def\H {{\cal H}}
\def\l {{\bf l}}

\def\XX{{\cal X}}
\def\YY{{\cal Y}}
\def\FF{{\cal F}}
\newcommand{\g}{{\bf g}}
\newcommand{\s}{{sup}}
\def\r {{\bf r}}
\def\t {{\bf t}}

\thispagestyle{plain}

\title{$H(X)$ vs. $H(f(X))$}
\author{
  \IEEEauthorblockN{Ferdinando Cicalese}
  \IEEEauthorblockA{
    Universit\`a di Verona,
    Verona, Italy\\
    Email: cclfdn22@univr.it } 
  \and
  \IEEEauthorblockN{Luisa Gargano
  \IEEEauthorblockA{
    Universit\`a di Salerno,
    Salerno, Italy\\
    Email: lgargano@unisa.it}
  \and
  \IEEEauthorblockN{Ugo Vaccaro}
  \IEEEauthorblockA{
    Universit\`a di Salerno,
    Salerno, Italy\\
    Email: uvaccaro@unisa.it} 
}
}
\maketitle
\begin{abstract}
It is well known that the entropy $H(X)$ of a finite random variable is always 
greater or equal to the entropy $H(f(X))$ of a function $f$ of $X$, with equality if and only if
 $f$ is one-to-one. In this paper, we give tights bounds on $H(f(X))$
when the function $f$ is not one-to-one, and we illustrate a few  scenarios where 
this matters. As an intermediate step towards our main result, we  
prove a lower  bound on the entropy of a probability distribution, when only a bound
on the ratio between  the maximum and the minimum probability 
is  known. Our lower bound 
improves previous results in the literature, and it could  
 find applications outside the present scenario.
\end{abstract}

\section{The Problem}
Let $\XX=\{x_1,  \ldots ,x_n\}$
be a  finite alphabet, and $X$ be any    random variable (r.v.)
taking values in $\XX$ according to the probability distribution
$\p=(p_1, p_2, \ldots , p_n)$, that is, such that $P\{X=x_i\}=p_i$, 
for $i=1, 2, \ldots , n$. A well known and widely used inequality (see \cite{CT}, Exercise 2.4),  states
that 
\begin{equation}\label{eq:HX>HfX}
H(f(X))\leq H(X), 
\end{equation}
where $f:\XX\to \YY$ is any function defined on $\XX$,  and $H(\cdot)$ denotes  the Shannon
entropy. Moreover, equality holds
in (\ref{eq:HX>HfX}) if and only  if
$f$ is one-to-one.
The main   purpose of this paper is to sharpen inequality (\ref{eq:HX>HfX}) by deriving tight bounds on $H(f(X))$ when 
$f$ \emph{is not} one-to-one. More precisely, given the r.v. $X$, an integer $2\leq m<n$, a set $\YY_m=\{y_1,  \ldots ,y_m\}$,
and the family of surjective functions  
$\FF_m=\{f| \;  f:\XX\to \YY_m,	\ |f(\XX)|=m \}$,
we want to compute the values
\begin{equation}\label{eq:maxandmin}
\max_{f\in \FF_m}H(f(X)) \qquad \mbox{and} \qquad \min_{f\in \FF_m}H(f(X)).
\end{equation}

\section{The Results}
For any probability distribution $\p=(p_1, p_2, \ldots , p_n)$, with $p_1\geq p_2, \ldots , \geq p_n\geq 0$, and integer 
$2\leq m<n$, let us define the probability distributions $R_m(\p)=(r_1, \ldots ,r_m)$ as follows:
{ if 
$p_1<1/m$ we set $R_m(\p)=(1/m, \ldots ,1/m)$, whereas if $p_1\geq 1/m$
we set  $R_m(\p)=(r_1, \ldots ,r_m)$, where  }
\begin{equation} \label{eq:definition-restriction}
r_i =  \begin{cases}
p_i & \hbox{ for } i = 1, \dots, i^* \cr
\left (\sum_{j=i^*+1}^n p_j\right )/{(m-i^*)} & \hbox{ for } i = i^*+1, \dots, m,
\end{cases}
\end{equation}
and  {$i^*$ is the maximum index $i$ such that 
$p_i\geq \frac{\sum_{j=i+1}^n p_j}{m-i}$.}
A somewhat similar operator was introduced in \cite{HY}.

\noindent
Additionally, we define the probability distributions $Q_m(\p)=(q_1, \ldots ,q_m)$ 
in the following way:
\begin{equation} \label{eq:Q}
q_i =  \begin{cases}
\sum_{k=1}^{n-m+1}p_k, \quad  & \hbox{ for } i = 1, \cr
p_{n-m+i}, & \hbox{ for } i=2, \ldots,  m.
\end{cases}
\end{equation}

The following Theorem provides  the results seeked in (\ref{eq:maxandmin}).
\begin{theorem}
For any   r.v.   $X$ 
taking values in the alphabet  $\XX=\{x_1, x_2, \ldots ,x_n\}$ according to the  probability distribution
$\p=(p_1, p_2, \ldots , p_n)$, and for any $2\leq m<n$, it holds that
\begin{equation}\label{max}
\max_{f\in \FF_m}H(f(X))\in \left[H(R_m(\p))-\alpha, H(R_m(\p))\right],
\end{equation}
where 
$\alpha=1-({1+\ln(\ln 2)})/{\ln 2}< 0.0861$, and
\begin{equation}\label{min}
\min_{f\in \FF_m}H(f(X))=H(Q_m(\p))\footnote{Here, with a slight 
abuse of notation, for a probability distribution $\A=(a_1, \ldots , a_t)$ we denote with 
$H(\A)=-\sum_ia_i\log a_i$ the entropy of a discrete r.v. distributed according to 
$\A$. Moreover, with $\log$ we denote the logarithm   in base 2, and with $\ln$ the natural
 logarithm in base $e$}.
\end{equation}
\end{theorem}
Therefore, the function $f\in \FF_m$ for which $H(f(X))$ is minimum maps all
the elements $x_1, \ldots , x_{n-m+1}\in \XX$ to a single element, and it is one-to-one
on the remaining elements $x_{n-m+2}, \ldots , x_n$.

Before proving Theorem 1 and discuss its consequences, we would like to notice that there are quite compelling
reasons why we are unable
to determine  the exact value of the maximum in (\ref{max}), and consequently, the form of 
the function $f\in \FF_m$ that attains the bound. Indeed, 
computing    the   value $\max_{f\in \FF_m}H(f(X))$ 
is an NP-hard problem. It is easy to understand the difficulty of the problem already
in the simple case $m=2$. To that purpose, consider any function $f\in \FF_2$, that is $f:\XX\to\YY_2=\{y_1, y_2\}$,
and let $X$ be any    r.v.
taking values in $\XX$ according to the probability distribution
$\p=(p_1, p_2, \ldots , p_n)$.
Let 
$z_1=\!\!\!\sum_{x\in \XX : f(x)=y_1}P\{X=x\}, \quad z_2=\!\!\!\sum_{x\in \XX : f(x)=y_2}P\{X=x\}.$
Then,  $H(f(X))=-z_1\log z_1-z_2\log z_2$,  and it  is maximal   
in correspondence
of a function $f\in \FF_2$ that makes the sums  $z_1$ and $z_2$ 
as much equal as possible. This is equivalent to the well known NP-hard problem  \textsc{Partition} 
on the instance $\{p_1, \ldots , p_n\}$ (see \cite{GJ})\footnote{In the full version of the paper we will show 
that the problem of computing    the   value $\max_{f\in \FF_m}H(f(X))$ 
is \emph{strongly} NP-hard}.
Since the function $f\in \FF_m$ 
for which $H(f(X))\geq H(R_m(\p))-\alpha$ can be efficiently
constructed, 
we have also the following important consequence of Theorem 1.
\begin{corollary}
There is a polynomial time algorithm to approximate the NP-hard problem of computing the value
$$\max_{f\in \FF_m}H(f(X)),$$
 with an \emph{additive} approximation factor of $\alpha\leq 0.0861$.
\end{corollary}
A key tool  for the proof of Theorem 1 is the following result, proved in 
the second part of Section \ref{proofs}.
\begin{theorem}\label{teo-H1}
	Let ${\bf p}=(p_1,p_2, \ldots, p_n)$ be a probability distribution such that  
$p_1\geq  p_2\geq \ldots \geq  p_n>0$. If $ p_1/p_n\leq \R$ then  
\begin{equation}\label{eq:rho}
H({\bf p})\geq \log n - \left(   \frac{\R \ln \R}{\R-1} -1-
                        \ln\frac{\R\ln \R}{\R-1}\right)\frac{1}{\ln 2}.
	\end{equation}											
\end{theorem}
Theorem \ref{teo-H1} improves on   several  papers (see \cite{S+} and 
references therein quoted), that have studied the   problem 
of estimating  $H(\p)$ when only a bound on the ratio $ p_1/p_n$ is 
known.\footnote{The bound in \cite{S+} has this form: if $p_1/p_n\leq 1+2(e^\epsilon-1)+2\sqrt{e^{2\epsilon}-e^\epsilon}$, then
$H(X)\geq \log n -\epsilon$. One can see that our bound (\ref{eq:rho}) is tighter.}
We believe  the result to be  of independent interest. 
For instance, it can also be used to improve existing bounds on the leaf-entropy  of parse trees generated by 
Tunstall  algorithm.

To prove our results, we  use ideas  and techniques  from Majorization Theory \cite{MO}, 
a mathematical framework that has been  proved to be very  much useful  in 
Information Theory (e.g., see \cite{CV,CV1,HY,HV} and references therein quoted).

\section{Some Applications}\label{app}
Besides its inherent naturalness, the problem of estimating the entropy $H(f(X))$
vs. $H(X)$
has several interesting applications. We highlight  some 
of them here, postponing a more complete discussion in 
the full version of the paper. 

In the area of clustering, one seeks a mapping $f$ (deterministic or stochastic)
from some data, generated  by a r.v. $X$ taking values in a set $\XX$, 
to ``clusters'' in $\YY$, where 
$|\YY|\ll |\XX|$. 
A widely employed  measure to appraise the goodness of a clustering algorithm is 
the information that the clusters retain towards the original data,
measured by the mutual information $I(X;f(X))$ (see \cite{F+,KMN} and references therein quoted). 
In general, one wants to choose $f$ such that  $|f(\XX)|$ is small
but $I(X;f(X))$ is large. The authors of  \cite{GA} (see also \cite{kur}) proved that, {given} the random
variable $X$, among all mappings  $f$ that maximizes $I(X;f(X))$ (under the
constraint that 
$|f(\XX)|$ is fixed) there is a maximizing function $f$ that 
is \emph{deterministic}. Since in the case of deterministic
functions it holds that   $I(X;f(X))=H(f(X))$, 
finding the clustering $f$ of $\XX$ (into a  fixed number $m$ of clusters) 
that maximizes the mutual
information $I(X;f(X))$ is \emph{equivalent} to our problem of finding the
function $f$ that attains the upper bound in (\ref{eq:maxandmin}).%
\footnote{In \cite{kur}
the authors consider the    problem of determining the function  $f$ that maximizes
$I(X;f(Y))$, where $X$ is the r.v. at the input of a DMC and $Y$ is the corresponding output. Our scenario could be
seen as the particular case when the DMC is noiseless.
However, the results in \cite{kur} do not imply ours  since the authors  give algorithms  only
for  binary input channels (i.e. $n=2$,  that makes the problem completely  trivial in our case). 
Instead, our results are relevant to those of \cite{kur}. For instance, we obtain  that the  general
maximization problem considered in \cite{kur} is NP-hard, a fact unnoticed in \cite{kur}.
}

Another scenario where our results directly find applications is the one considered
in \cite{V12}. There, the author considers the problem
of best approximating a probability distribution 
$\p=(p_1, \ldots, p_n)$ with a shorter  one
$\q^*=(q^*_1, \ldots , q^*_m)$, $m\leq n$.
The criterion with which one chooses $\q^*$, given $\p$,
is the following. Given  
$\p=(p_1, \ldots, p_n)$ 
and $\q=(q_1, \ldots , q_m)$, 
define the 
quantity $\D(\p,\q) $ as $2W(\p,\q)-H(\p)-H(\q)$,
where $W(\p,\q)$ is the \emph{minimum} entropy of a bivariate
probability distribution that has $\p$ and $\q$ as marginals.
Then, 
the ``best'' approximation $\q^*$ of $\p$ is chosen as the
probability distributions $\q^*$ with $m$ components that \emph{minimizes} $\D(\p,\q)$,
over all $\q=(q_1, \ldots , q_m)$.
The author of \cite{V12} 
shows  that $\q^*$ can  be characterized in the following way. 
Given $\p=(p_1, \ldots , p_n)$, 
call  $\q=(q_1, \ldots , q_m)$ an \emph{aggregation} 
of $\p$ into $m$ components if there is a partition of $\{1, \ldots , n\}$ into disjoint sets $I_1, \ldots , I_m$
such that $q_k=\sum_{i\in I_k}p_i$, for $k=1, \ldots m$.
In \cite{V12} it is proved that  the vector $\q^*$ that best 
approximate $\p$ (according to $\D$) is the aggregation of $\p$ 
into $m$ components of \emph{maximum entropy}.
Since \emph{any} aggregation $\q$ of $\p$ can 
be   seen as the distribution of the r.v. 
 $f(X)$, where $f$ is  some appropriate function
 and  $X$ is  a r.v. distributed according to
$\p$ (and, vice versa, any deterministic $f$ \emph{gives} a r.v. 
$f(X)$ whose distribution is an aggregation of the distribution of $X$), 
one gets that  the problem of computing the ``best'' approximation
$\q^*$ of $\p$ is NP-hard. 
The bound (\ref{max})  
allows us to provide an approximation algorithm 
to construct a probability distribution $\overline{\q}=(\overline{q}_1, \ldots , \overline{q}_m)$
such that $\D(\p,\overline{\q})\leq \D(\p,\q^*)+0.0861$, improving 
on  
\cite{CGV}, where an approximation algorithm
for the same problem with an additive error of $1$ was provided. 

There are  other problems that can be cast in our scenario.
For instance, Baez \emph{et al}. \cite{Baez} give an axiomatic
characterization of the Shannon entropy in
terms of  \emph{information loss}. Stripping
away the Category Theory language of \cite{Baez}, the information loss of a r.v. $X$ amounts
to the difference $H(X)-H(f(X))$, where $f$ is any deterministic function.
Our Theorem 1  allows  to quantify the extreme value of the information
loss of a r.v., when the support of $f(X)$ is known.

There is also  a vast literature (see \cite{Ma}, Section 3.3,  and references therein quoted)
 studying  the ``\emph{leakage of a program $P$  [...]  defined as the
(Shannon) entropy of the partition} $\Pi(P)$''  \cite{Ma}. One can easily see that their ``leakage''
is  the same as   the entropy $H(f(X))$, where 
$X$ is the r.v. modeling the program input, and $f$ is the
function describing the input-output relation of the program $P$.
In Section 8 of the same paper the authors study the problem of 
maximizing or minimizing the leakage, in the case the program $P$ is 
stochastic, using standard techniques based on Lagrange multipliers.
They do not consider the (harder) case of deterministic programs (i.e.,
deterministic $f$'s) and our results are likely to be relevant in 
that context.

Finally, we remark that our problem can also be seen as a problem 
of quantizing the alphabet  of a  discrete source  into a smaller one (e.g., \cite{ME}),
and   the goal  is to maximize the mutual information between  
the original source and the quantized one.
\section{The Proofs}\label{proofs}
 We first recall the important concept of
\emph{majorization} among probability distributions.
\begin{definition}\label{defmaj} {\rm \cite{MO}}
Given two probability distributions
$\A=(a_1, \ldots ,a_n)$ and $\B=(b_1, \ldots , b_n)$ with $a_1\geq \ldots \geq a_n\geq 0$ and 
$b_1\geq \ldots \geq b_n\geq 0$, we say that $\A$ is 
{\em majorized} by $\B$, and write  $\A \preceq \B$,
if and only if
$$\sum_{k=1}^i a_k\leq \sum_{k=1}^i b_k, \quad\mbox{\rm for all }\  i=1,\ldots , n.$$
\end{definition}
Without loss of generality we assume that \emph{all} the probabilities distributions 
we deal with have been
ordered in non-increasing order. We also  use the majorization 
relationship between vectors of unequal lenghts, by properly padding the shorter
one with the appropriate number of $0$'s at the end.

Consider   an arbitrary function $f:\XX\to\YY$,
$f\in \FF_m$.
Any r.v. $X$ taking values in $\XX=\{x_1, \ldots , x_n\}$,  according to the probability distribution
$\p=(p_1, \ldots, p_n)$,  and the function $f$ naturally induce 
a  r.v. $f(X)$, taking values in $\YY=\{y_1, \ldots , y_m\}$
according to the 
probability distribution 
whose values are given by the expressions 
\begin{equation}\label{eq:defy}
\forall y_j\in \YY \qquad P\{f(X)=y_j\}=\sum_{x\in \XX:f(x)=y_j}P\{X=x\}.
\end{equation}
Let $\z=(z_1, \ldots , z_m)$ be the  vector containing the values  
$z_1=P\{f(X)=y_1\}, \ldots , z_m=P\{f(X)=y_m\}$ ordered in non-increasing fashion.
For convenience, we state the following self-evident fact about the
relationships between $\z$ and $\p$.
\begin{claim}
There is a partition of $\{1, \ldots , n\}$ into disjoint sets $I_1, \ldots , I_m$
such that $z_j=\sum_{i\in I_j}p_i$, for $j=1, \ldots m$. 
\end{claim}
\noindent
Therefore,  $\z$ is an \emph{aggregation} of $\p$. 
Given a r.v. $X$ distributed according to $\p$, and \emph{any} $f\in \FF_m$,  by simply applying 
the definition
of majorization one can see that the (ordered) probability distribution 
of the r.v. $f(X)$ is majorized by  $Q_m(\p)=(q_1, \ldots ,q_m)$,  as defined in 
(\ref{eq:Q}).
Therefore, by invoking the Schur concavity  
of the entropy function $H$ (see \cite{MO}, p. 101 for the statement,  and \cite{HV} for an improvement), saying that 
 $H(\A)\geq H(\B)$ whenever $\A\preceq \B$, we get that 
 $H(f(X))\geq H(Q_m(\p))$. From this, the equality (\ref{min}) immediately  follows.
 
We  need the following two simple  results, but important to us, stated and proved in \cite{CGV}
with a  different terminology.
\begin{lemma}{\rm\cite{CGV}}
For $\p$ and $\z$ as above, it holds that 
$\p\preceq \z.$
\end{lemma}
In other words, {for any} r.v. $X$ {and function} $f$, the probability distribution of $f(X)$
\emph{always} 
{majorizes}  that  of $X$.
\begin{lemma}{\rm\cite{CGV}}
For \emph{any} $m$, $2\leq m<n$,  and probability distribution $\A=(a_1, \ldots , a_m)$ such that $\p\preceq \A$, 
it holds that 
\begin{equation}\label{p<R<a} 
R_m(\p)\preceq \A, 
\end{equation}
where $R_m(\p)$ is the probability distribution defined in {\rm (\ref{eq:definition-restriction}). }
\end{lemma}
 From Lemmas 1 and 2, and by applying  the Schur concavity  
of the entropy function $H$, we get the following result.
\begin{corollary}
For any  r.v. $X$ 
taking values in $\XX$ according to a probability distribution
$\p$, and 
for \emph{any} $f\in \FF_m$,  
it holds that
\begin{equation}\label{HF<HR}
H(f(X))\leq H(R_m(\p)).
\end{equation}
\end{corollary}
\noindent
Above corollary implies that 
$$\max_{f\in \FF_m}H(f(X))\leq H(R_m(\p)).$$
Therefore, to complete the proof of Theorem 1 we need to show that we can construct a  
function $f\in \FF_m$ such that 
\begin{equation}\label{need}
H(f(X))\geq H(R_m(\p))-\left(1-\frac{1+\ln(\ln 2)}{\ln 2}\right),
\end{equation}
or, equivalently, that we can construct    an {aggregation} of $\p$  into $m$ components, 
whose entropy is at least $H(R_m(\p))- \left(1-\frac{1+\ln(\ln 2)}{\ln 2}\right).$
We prove this fact  in  the following lemma.
\begin{lemma} \label{lemma:huffman-prob}
For any $\p=(p_1, \ldots , p_n)$ and $2\leq m<n$, we can construct {an aggregation} ${\bf q}=(q_1, \ldots , q_m)$
of $\p$
 such that
$$H({\bf q}) \geq H(R_m(\p)) -  \left(1-\frac{1+\ln(\ln 2)}{\ln 2})\right).$$
\end{lemma}
\begin{IEEEproof}
We will assemble the aggregation ${\bf q}$ through the Huffman algorithm.
We first make the following observation. To the  purposes of this paper, each  \emph{step} of the 
Huffman algorithm consists in merging the two smallest 
element  $x$ and $y$ of the current probability distribution, 
deleting $x$ and $y$  and substituting them with the single element $x+y$,
and \emph{reordering} the new probability distribution from 
the largest element to the smallest (ties are arbitrarily broken). 
Immediately after the step in which $x$ and $y$ are merged,  \emph{each} element $z$ in the new and reduced
probability distribution that finds itself positioned  at the ``right'' of 
$x+y$ (if there is such a $z$)
 has a value  that satisfies  $(x+y)\leq 2z$ (since, by choice, $x,y\leq z$).
Let ${\bf q} = (q_1, \dots, q_m)$  be the ordered probability 
distribution obtained by executing \emph{exactly} $n-m$ steps 
of the Huffman algorithm, starting from the distribution ${\bf p}$.
Denote by  $i_q$  the maximum index $i$ 
such that for each $j = 1, \dots ,i_q$ the component $q_j$ \emph{has not} been produced by 
a merge operation of the Huffman algorithm. In other word, 
$i_q$ is the maximum index $i$ 
such that for each $j = 1, \dots ,i_q$ it holds that $q_j=p_j$. Notice that we allow $i_q$ to be equal to $0$.
Therefore $q_{i_q+1}$ has been produced by a merge operation. 
At the   step in which the value $q_{i_q+1}$ was created,  it holds that $q_{i_q+1}\leq 2z$,
for any $z$ at the ``right'' of  $q_{i_q+1}$. At later steps, the inequality 
$q_{i_q+1}\leq 2z$ still holds, since elements
at the right of $q_{i_q+1}$ could have  only increased their values.

Let $S = \sum_{k=i_q+1}^mq_k$ be the sum of the last (smallest) $m-i_q$ components of ${\bf q}$. 
The vector  
${\bf q}' = (q_{i_q+1}/S,  \dots q_{m}/S)$
is a probability distribution such that the  ratio between its
 largest and its smallest 
component 
is upper bounded by 2.
By Theorem \ref{teo-H1}, with $\rho=2$,  it   follows that 
\begin{equation} \label{equation:DG}
H({\bf q}') \geq \log(m-i_q) - \alpha,
\end{equation}
where $\alpha\leq \left(1-\frac{1+\ln(\ln 2)}{\ln 2}\right)< 0.0861$.
Therefore, we have 
\begin{eqnarray*}
H({\bf q}) &=& \sum_{j=1}^{i_q} q_j \log \frac{1}{q_j} + \sum_{j=i_q+1}^{m} q_j \log \frac{1}{q_j}  \label{eq:h-1st}\\
&=& \sum_{j=1}^{i_q} q_j \log \frac{1}{q_j} - S \log S + S \sum_{j=i_q+1}^{m} \frac{q_j}{S} \log \frac{S}{q_j}  \\
&=& \sum_{j=1}^{i_q} q_j \log \frac{1}{q_j} - S \log S + S H({\bf q}')\\
&\geq& \sum_{j=1}^{i_q} q_j \log \frac{1}{q_j} - S \log S + S (\log(m-i_q) - \alpha) \\
&=& \sum_{j=1}^{i_q} q_j \log \frac{1}{q_j} + \!\!S \log\frac{m-i_q}{S} -  \alpha S\\
&=&  \sum_{j=1}^{i_q} q_j \log \frac{1}{q_j}\! + \!\!\!\sum_{j=i_q+1}^m \frac{S}{m-i_q} \log\frac{m-i_q}{S}  - \alpha S  \\
&\geq&  \sum_{j=1}^{i_q} q_j \log \frac{1}{q_j} + \sum_{j=i_q+1}^m \frac{S}{m-i_q} \log\frac{m-i_q}{S}  - \alpha \\
&=& H\Bigl(q_1, q_2, \dots, q_{i_q}, \frac{S}{m-i_q},  \dots, \frac{S}{m-i_q}\Bigr) - \alpha. \label{eq:h-last} 
\end{eqnarray*}

Let ${\bf q}^* = (q_1, q_2, \dots, q_{i_q}, \frac{S}{m-i_q}, \frac{S}{m-i_q}, \dots, \frac{S}{m-i_q}),$ 
and observe that  
${\bf q}^*$ coincides with ${\bf p}$ in the first $i_q$ components, as it does $\q$.
What we have shown is that 
\begin{equation} \label{h-h*}
H({\bf q}) \geq H({\bf{q}}^*) - \alpha.
\end{equation}

We now observe that $i_q \leq i^*$, where $i^*$ is the index that intervenes in the definition of  our  operator
$R(\p)$ (see (\ref{eq:definition-restriction})). 
In fact, by the definition of $\q$ one has  $q_{i_q} \geq q_{i_q+1} \geq \cdots \geq q_m$, that  also implies  
\begin{equation}\label{eq:hh}
\frac{\sum_{j=i_q+1}^m q_j}{m} \leq q_{i_q+1} \leq q_{i_q} = p_{i_q}.
\end{equation}
Moreover, since the first $i_q$ components of ${\bf q}$ are the same as in ${\bf p}$, we  also have
$\sum_{j = i_{q}+1}^m q_j = \sum_{i_q+1}^n p_j$. This, together with   relation (\ref{eq:hh}), implies
\begin{equation}\label{eq:new}
\frac{\sum_{j=i_q+1}^n p_j}{m} \leq p_{i_q}.
\end{equation}
Equation (\ref{eq:new}) clearly implies $i_q \leq i^*$ since $i^*$ is
by definition,  the maximum index $i$ such that $\sum_{j=i+1}^n p_j \geq (n-i) p_i.$
From the just proved inequality  $i^* \geq i_q$, we have also
\begin{equation} \label{h:inequality}
{\bf q}^* \preceq R({\bf p}). 
\end{equation}
Using (\ref{h-h*}), (\ref{h:inequality}),  and the Schur concavity of the entropy function, we get 
$$H({\bf q})\geq H({\bf q}^*)-\alpha\geq H(R({\bf p}))-\alpha,$$
thus completing the proof of the Lemma (and  of Theorem 1).
\end{IEEEproof}

We now prove  Theorem \ref{teo-H1}.
Again, we  use  tools from   majorization theory.
Consider an arbitrary  probability distribution  ${\bf p}=(p_1,p_2, \ldots, p_n)$ with  
$p_1\geq  p_2\geq \ldots \geq  p_n>0$ and  $p_1/p_n\leq \R$.
Let us define  the  probability distribution 
\begin{eqnarray}\label{zeta}
\bz_\R({\bf p})=(z_1,\ldots,  z_n)&&\\
              =(\underbrace{\R p_n,\ldots, \R p_n}_{i \ \mbox{\scriptsize times}},&&
							                                     \hspace{-0.7truecm} 1-(n+i\R-i-1)p_n,p_n, \ldots, p_n),\nonumber
\end{eqnarray}
where 
 $i=\left\lfloor{(1-np_n)}/{p_n(\R-1)}\right\rfloor$. 
It is easy to verify that  $p_n\leq 1-(n+i(\R-1)-1)x\leq \R p_n$.
\begin{lemma}\label{primo}
Let ${\bf p}=(p_1,p_2, \ldots, p_n)$ with  
$p_1\geq  p_2\geq \ldots \geq  p_n>0$  be any probability distribution  with  $p_1/p_n\leq\R$.
The  probability distribution 
 $\bz_\R({\bf p})$
 satisfies
	${\bf p}\preceq \bz_\R({\bf p}).$
\end{lemma}
\begin{IEEEproof}
For any $j\leq i$, it holds that
$$p_1+\ldots + p_j\leq j\, p_1\leq j (\R p_n)=z_1+\ldots +z_j.$$
Consider now some  $j\geq i +1$ and  assume by contradiction that 
$p_1+\ldots + p_j >z_1+\ldots+z_j$.
It follows that 
$p_{j+1}+\ldots+p_n<z_{j+1}+\ldots+z_n=(n-j)p_n$. As a consequence 
we get the contradiction  $p_n\leq (p_{j+1}+\ldots+p_n)/(n-j)<p_n$.
\end{IEEEproof}

\medskip
Lemma  \ref{primo} and the Schur concavity of the entropy imply  that $H(\bp)\geq H(\bz_\R({\bf p}))$. 
	We  can therefore prove Theorem 2  by showing    the appropriate  upper  bound on $\log n -H(\bz_\R({\bf p}))$.
\begin{lemma}
	It holds that 
	$$\log n -H(\bz_\R({\bf p}))\leq \left(\frac{\R\ln \R}{\R-1} -1- \ln \frac{\R\ln \R}{\R-1}\right)\frac{1}{\ln 2}.$$
	
\end{lemma}
\begin{IEEEproof}
Consider  the class of  probability distributions of the form
$$\bz_\R(x,i)=(\R x,\ldots, \R x,1-(n+i(\R-1)-1)x,x, \ldots, x),$$
having  the first $i$ components equal to  ${\R x}$ and the last $n-i-1$ equal to $x$, for  
suitable  $0\leq x\leq 1/\rho$, and $i\geq 0$ 
such that
 \begin{equation}\label{xin}
1-(n+i(\R-1)-1)x\in[x,\R x).
\end{equation}
Clearly, for $x=p_n$ and $i=\left\lfloor{(1-np_n)}/{p_n(\R-1)}\right\rfloor$ one has    $\bz_\R({\bf p})=\bz_\R(x,i)$, and we can prove the lemma by 
upper bounding the 
maximum (over all  $x$ and $i$)
of  $\log n -H(\bz_\R(x,i))$.
Let
\begin{align*}
f(x,&i)= \log n -H(\bz_\R((x,i)) =\log n + i (\R x\log (\R x))\\
 +&   (1-(n+i(\R-1)-1)x)\log(1-(n+i(\R-1)-1)x) \\
& \ \qquad\qquad\qquad \qquad\qquad\qquad\qquad  + (n-i-1)x \log x.
\end{align*}
From (\ref{xin}), for any value of  $i\in \{1,\ldots, n-2\}$, one has that
\begin{equation*}
x\in \left(\frac{1}{n+(i+1)(\R-1)},\frac{1}{n+i(\R-1)}\right]
\end{equation*}
Set $A=n+i(\R-1)-1$. We have
\begin{align*}
f(x,i)=&\log n +i\R x \log(\R x)\\
        & -(1-Ax) \log (1-Ax)+(n-i-1)x\log x,\\
\frac{d}{d x}f(x,i)=&i\R\log \R + (i\R -A +n-i-1)\log e \\
                     &+ (i\R +n-i-1)\log x -A\log(1-Ax)\\
										=&i\R\log \R+ A\log x-A\log(1-Ax),\\
\frac{d^2}{d x^2}f(x,i)=& \Bigl(\frac{A}{x}
                          +\frac{A^2}{1-Ax}\Bigr ) \log e.
													\end{align*}
Since 
$\frac{d^2}{d x^2}f(x,i)\geq 0$ for 
any 
$x\in \left(\frac{1}{n+(i+1)(\R-1)},\frac{1}{n+i(\R-1)}\right]$, 
the function is $\cup$-convex in this interval, and it is upper bounded by 
the maximum between the two extrema values $f(1/(n+(i+1)(\R-1)),i)$ and $f(1/(n+i(\R-1)),i)$.
Therefore, we can upper bound $f(x,i)$ by  the maximum value among 
\begin{align*}
f(1/(n+i(\R-1)),i)=&\log n+ \frac{i\R}{n+i(\R-1)}\log \R \\
                     &+\log \frac{1}{n+i(\R-1)},
 \end{align*}
for $i=1,\ldots, n-1$. We now interpret  $i$ as a continuous variable, and 
we differentiate   $\log n+ \frac{i\R}{n+i(\R-1)}\log \R +\log \frac{1}{n+i(\R-1)}$
with respect to $i$.  We get  
\begin{align*}
\frac{d}{d i} & \left(\log n+ \frac{i\R}{n+i(\R-1)}\log \R +\log \frac{1}{n+i(\R-1)}\right)\\
               &=
    \frac{n(\R\log \R-(\R-1)\log e)-i(\R-1)^2\log e}{(n+i(\R-1))^2},
    \end{align*}
that is positive if and only if 
$i\leq \frac{n}{\R-1}\left(\frac{\R\ln \R}{\R-1}-1\right).$
Therefore, the desired upper bound on $f(x,i)$ can be obtained by computing the value of 
$f(\overline{x},\overline{\imath})$, where 
$\overline{\imath}=\frac{n}{\R-1}\left(\frac{\R\ln \R}{\R-1}-1\right)$ and 
$\overline{x}=\frac{1}{n+\overline{\imath}(\R-1)}$. The  value of 
$f(\overline{x},\overline{\imath})$ turns out to be equal to
\begin{align*}
\log n &+
\frac{\frac{n}{\R-1}\left(\frac{\R\ln \R}{\R-1}-1\right)\R\log \R}{n+ n\left(\frac{\R\ln \R}{\R-1}-1\right)}
-\log \left(n+ n\!\left(\frac{\R\ln \R}{\R-1}-1\right)\!\right)\\
&= \frac{\R\log\R(\R\ln\R-\R+1)}{(\R-1)\R\ln \R}-\log\left(\frac{\R\ln\R}{\R-1}\right)\\
&=\frac{\R\ln\R-(\R-1)}{(\R-1)\ln 2} -\log\left(\frac{\R\ln\R}{\R-1}\right) \\
&=\left(\frac{\R\ln \R}{\R-1} -1- \ln\frac{\R\ln \R}{\R-1}\right)\frac{1}{\ln 2}.
\end{align*}
\end{IEEEproof} 
We  conclude the paper by showing how  Theorems 1 and 2 allow us to design an approximation algorithm 
for the second  problem mentioned in Section \ref{app}, that is, the 
problem of 
constructing  a probability distribution $\overline{\q}=(\overline{q}_1, \ldots , \overline{q}_m)$
such that $\D(\p,\overline{\q})\leq \D(\p,\q^*)+0.0861$. Our algorithm improves
on the result presented in \cite{CGV}, where an approximation algorithm
for the same problem with an additive error of $1$ was provided. 

\smallskip
Let ${\bf q}$ be  the probability distribution
constructed in Lemma 3 and let us recall that the first $i_q$ components of ${\bf q}$ coincide with the first $i_q$ components of ${\bf p}$. 
In addition, for each $i = i_q+1, \dots, m,$ there is a set $I_i \subseteq \{i_q+1, \dots, n\}$ such that
$q_i = \sum_{k \in I_i} p_k$ and the 
$I_i$'s form a partition of $\{i_q+1, \dots, n\},$
(i.e., ${\bf q}$ is an aggregation of $\p$ into $m$ components).
 
We now  build a bivariate probability distribution  ${\M}_q=[m_{ij}]$,
having  $\p$ and ${\bf q}$ as marginals,    as follows:
\begin{itemize}
\item in the first $i_q$ rows and columns, the matrix ${\M}_q$   has non-zero components  only  on the 
diagonal, namely $m_{j\,j} = p_j = q_j$ and $m_{i\, j} = 0$ for any $i, j \leq i_q$ such that $i \neq j$;
\item for each row $i = i_q+1, \dots, m$ the only non-zero elements are the ones in the columns 
corresponding to elements of $I_i$ and precisely, 
for each $j \in I_i$ we set $m_{i \, j} = p_j.$
\end{itemize}

It is not hard to see that $\M_q$ has  ${\bf p}$ and ${\bf q}$ as marginals.
Moreover we have that $H(\M_q) = H({\bf p})$ since by construction the only non-zero components of $\M_q$
coincide with the set of components of ${\bf p}.$ 
Let ${\cal C}(\p, \q)$ be the set of all bivariate probability distribution  
having  $\p$ and ${\bf q}$ as marginals. Recall that
$\alpha=1-({1+\ln(\ln 2)})/{\ln 2}< 0.0861$.
We have that
\begin{eqnarray} \label{eq:V-distance}
\D(\p, \q)&=& \min_{\N \in {\cal C}(\p, \q)} 2H(\N) - H(\p)  - H({\bf q})   \label{eq:V-2} \\
&\leq& 2 H(\M_q) - H(\p) - H({\bf q}) \label{eq:V-3} \\
&=& H({\bf p}) - H({\bf q}) \label{eq:V-3-1}\\
&\leq& H({\bf p}) - H(R_m(\p)) + \alpha \label{eq:V-5}\\
&\leq& H({\bf p}) - H({\bf q}^*) + \alpha \label{eq:V-5new}\\
&\leq& \D({\bf p}, {\bf q}^*) + \alpha \label{eq:V-6}
\end{eqnarray}
where
(\ref{eq:V-2}) is  the definition of $\D(\p, \q)$;
(\ref{eq:V-3}) follows from (\ref{eq:V-2})  since  $\M_q \in {\cal C}(\p, \q)$;
(\ref{eq:V-3-1}) follows from (\ref{eq:V-3})  because of $H(\M) = H(\p)$; 
(\ref{eq:V-5}) follows from  Lemma \ref{lemma:huffman-prob};
(\ref{eq:V-5new}) follows from  (\ref{eq:V-5}), the known fact 
that ${\bf q}^*$ is an aggregation of $\p$ (see \cite{V12}) and  Lemmas 1 and 2.
Finally, the general inequality  $H({\bf a}) - H({\bf b})\leq \D({\bf a}, {\bf b})$
is  formula (48) in \cite{KSS15}.

\end{document}